\documentclass[onecolumn,a4paper]{article}%
\usepackage{amsfonts}
\usepackage{amsmath}
\usepackage{amssymb}
\usepackage{graphicx}%
\usepackage{color}
\usepackage{hyperref}
\usepackage{cleveref}
\usepackage{subfig}

\newcommand{\bv}{\boldsymbol{v}}

\newcommand{\bh}{\boldsymbol{h}}

\newcommand{\btheta}{\boldsymbol{\theta}}

\begin{document}

\title{p-Adic Statistical Field Theory and Convolutional Deep Boltzmann Machines}
\author{W. A. Z\'{u}\~{n}iga-Galindo\thanks{The author was partially supported by the Lokenath Debnath Endowed Professorship.} \\ University of Texas Rio Grande Valley\\School of Mathematical and Statistical Sciences\\One West University Blvd.,\\Brownsville, TX 78520, United States.\\ wilson.zunigagalindo@utrgv.edu 
\and 
Cuiyu He\\Oklahoma State University, Department of Mathematics\\MSCS 425, Stillwater, OK, United States.\\ cuiyu.he@okstate.edu
\and 
B. A. Zambrano-Luna\\University of Texas Rio Grande Valley\\School of Mathematical and Statistical Sciences\\One West University Blvd.,\\Brownsville, TX 78520, United States.\\ brian.zambrano@utrgv.edu}
\maketitle

\begin{abstract}
Understanding how deep learning architectures work  is a central scientific
problem. Recently, a correspondence
between neural networks (NNs) and Euclidean quantum field theories (QFTs) has been proposed. This work investigates this correspondence in the framework of $p$-adic statistical
field theories (SFTs) and neural networks (NNs). In this case, the fields are
real-valued functions defined on an infinite regular rooted tree with valence
$p$, a fixed prime number. This infinite tree provides the topology for a
continuous deep Boltzmann machine (DBM),  which is  identified with a statistical
field theory (SFT) on  this  infinite tree.
In the $p$-adic framework, there is a natural method to discretize SFTs. Each discrete SFT corresponds to a Boltzmann machine (BM) with a tree-like topology. This method allows us to recover the standard DBMs and gives new convolutional DBMs. The new networks use $O(N)$ parameters while the classical ones use $O(N^2)$ parameters.
\end{abstract}

\section{Introduction}

The deep neural networks have been successfully applied to various
 tasks, including self-driving cars, natural language processing, and visual
recognition, among many others, \cite{Lecun-Benigio}-\cite{Bahri et al}. There
is consensus about the need of developing a theoretical framework to
understand how deep learning architectures work. Recently, physicists have
proposed the existence of a correspondence between neural networks (NNs) and
quantum field theories (QFTs), more precisely statistical field theory (SFT), see
\cite{Buice}-\cite{Zuniga1}, and the references therein. This correspondence
takes different forms depending on the architecture of the networks involved.

In \cite{Zuniga1}, the study of the above-mentioned correspondence was
initiated in the framework of the non-Archimedean statistical field theory
(SFT).  
In this case, the background space (the set of real numbers) is replaced
by the set of $p$-adic numbers, where $p$ is a fixed prime number. 
The $p$-adics are organized in a tree-like structure; this feature facilitates the
description of hierarchical architectures. In \cite{Zuniga1}, 
a $p$-adic version of the convolutional deep Boltzmann machines is introduced where only binary data is considered and with no implementation. By adapting the mathematical techniques introduced by Le Roux and Benigio in \cite{Le roux et al 1}, the author shows that these machines are universal approximators for binary data tasks. 

In this article, we continue discussing the correspondence between STFs and NNs, in the
$p$-adic framework. Compare with \cite{Zuniga1} here we consider more general architectures and data types. We note that dealing with general data is challenging both in theory and implementation practice.
We argue that $p$-adic analysis still provides the right
framework to understand the dynamics of NNs with large tree-like hierarchical
architectures. In our approach, a NN corresponds to the discretization of a
$p$-adic STF. The discretization process is carried out in a rigorous and
general way. Moreover, such discretization allows us to obtain many recently developed deep BM.
 For instance, the NNs constructed in \cite{Batchits et al} are a
particular case of the ones introduced here. 
We also discuss the implementation of a class of $p$-adic convolutional networks
and obtain desired results on a feature detection task based on hand-writing images of  decimal digits.

The main novelty of our $p$-adic convolutional DBMs is that they use significantly fewer parameters than the conventional ones. A detailed discussion is given in Section \ref{Section5}.  We note that the connections between $p$-adic numbers and neural networks have been considered before. Neural networks whose states are $p$-adic numbers were studied in \cite{Albeverio-Khrennikov-Tirozzi, Krennikov-tirozzi}. These models are completely different from the ones considered here. These ideas have been used to develop non-Archimedean models of brain activity and mental processes \cite{Khrenikov2}. In \cite{ZZ1, ZZ2}, p-adic versions of the cellular neural networks were studied. These models involved abstract evolution equations.

\section{\label{Section1}$p$-Adic Numbers}
In this section, we introduce basic concepts for the $p$-adic numbers. For more detailed information, refer \Cref{Section6} (Appendix A).

From now on, $p$ denotes a fixed prime number. Any non-zero $p-$adic number
$x$ has a unique expansion of the form%
\[
x=x_{-k}p^{-k}+x_{-k+1}p^{-k+1}+\ldots+x_{0}+x_{1}p+\ldots,\text{ }%
\]
with $x_{-k}\neq0$, where $k$ is an integer, and the $x_{j}$s \ are numbers
from the set $\left\{  0,1,\ldots,p-1\right\}  $. The set of all possible numbers of such form constitutes the field of $p$-adic numbers $\mathbb{Q}_{p}$. There
are natural field operations, sum, and multiplication, on $p$-adic numbers,
see, e.g., \cite{Koblitz}. There is also a natural norm in $\mathbb{Q}_{p}$
defined as $\left\vert x\right\vert _{p}=p^{k}$ where $k$ depends on $x$, for a nonzero $p$-adic number
$x$.%

\begin{figure}
[ptb]
\begin{center}
\includegraphics[
width=\textwidth
]%
{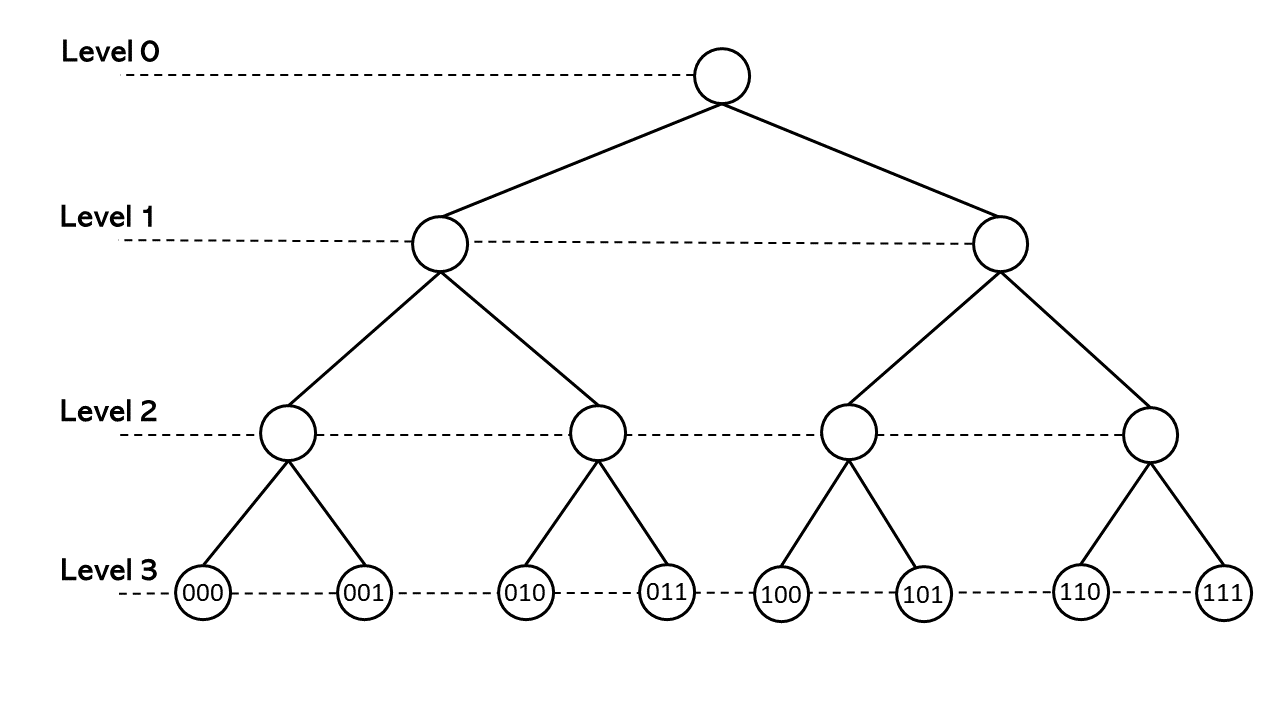}%
\caption{The rooted tree associated with the group $\mathbb{Z}_{2}%
/2^{3}\mathbb{Z}_{2}$. The elements of $\mathbb{Z}_{2}/2^{3}\mathbb{Z}_{2}$
have the form $i=i_{0}+i_{1}2+i_{2}2^{2}$,$\;i_{0}$, $i_{1}$, $i_{2}%
\in\{0,1\}$. The distance satisfies $-\log_{2}\left\vert i-j\right\vert
_{2}=$level of \noindent the first common ancestor of $i$, $j$.}%
\label{Figure 1}%
\end{center}
\end{figure}

The field of $p$-adic numbers with the distance induced by $\left\vert
\cdot\right\vert_{p}$ is a complete ultrametric space. The ultrametric
property refers to the fact that $\left\vert x-y\right\vert _{p}\leq
\max\left\{  \left\vert x-z\right\vert _{p},\left\vert z-y\right\vert
_{p}\right\}  $ for any $x$, $y$, $z$ in $\mathbb{Q}_{p}$. In this article, we
work with $p$-adic integers, which $p$-adic numbers satisfying $-k\geq0$. All
such $p$-adic integers constitute the unit ball $\mathbb{Z}_{p}$. The unit ball is closed under addition and multiplication, so it is a commutative ring. Along this article,
we work mainly with locally constant functions supported in the unit ball,
i.e.,  with functions of type $\varphi:\mathbb{Z}_{p}\rightarrow\mathbb{R}$,
such that $\varphi\left(  a+x\right)  =\varphi\left(  a\right)  $ for all $x$ in $\mathbb{Z}_p$. The simplest example of such function is the chararcterisstic function $1_{%
\mathbb{Z}_{p}}\left( x\right) $ of the unit ball $\mathbb{Z}_{p}$: $1_{%
\mathbb{Z}_{p}}\left( x\right) =1$ if $\left\vert x\right\vert _{p}\leq 1$,
otherwise $1_{\mathbb{Z}_{p}}\left( x\right) =0$. To check that $1_{\mathbb{Z%
}_{p}}\left( a+x\right) =1_{\mathbb{Z}_{p}}\left( a\right) $,  we use that $%
\mathbb{Z}_{p}$ is closed under addition.
 If
$\left\vert x\right\vert _{p}\leq p^{-l}$, where the integer $l$ is fixed and
independent of $a$. We denote by $\mathcal{D}(\mathbb{Z}_{p})$ the real vector
space of test functions supported in the unit ball. There is a natural
integration theory so that $\int_{\mathbb{Z}_{p}}\varphi\left(  x\right)  dx$ 
gives a well-defined real number. The measure $dx$ is the so-called Haar measure of
$\mathbb{Q}_{p}$.   Further details are given in Section \ref{Sec: Haar measure}.

Since the $p$-adic numbers are infinite series, any computational implementation involving these numbers requires a truncation process: $x\longmapsto x_{0}+x_{1}p+\ldots+x_{l-1}p^{l-1}$, $l\geq1$. The set of all
truncated integers mod $p^{l}$ is denote as $G_{l}=\mathbb{Z}_{p}%
/p^{l}\mathbb{Z}_{p}$. This set can be represented as a rooted tree with $l$ levels; see
\Cref{Figure 1}.

\begin{figure}
[ptb]
\begin{center}
\includegraphics[
width=0.5\textwidth]%
{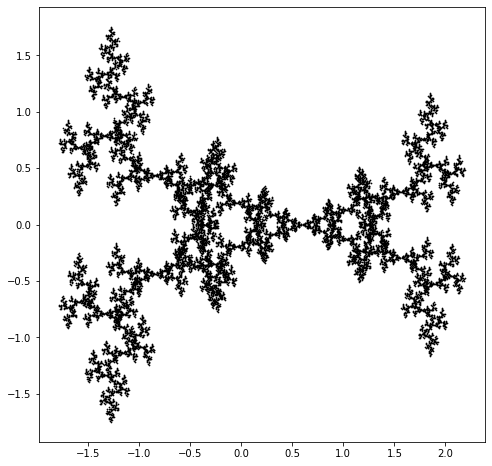}%
\includegraphics[
width=0.5\textwidth]%
{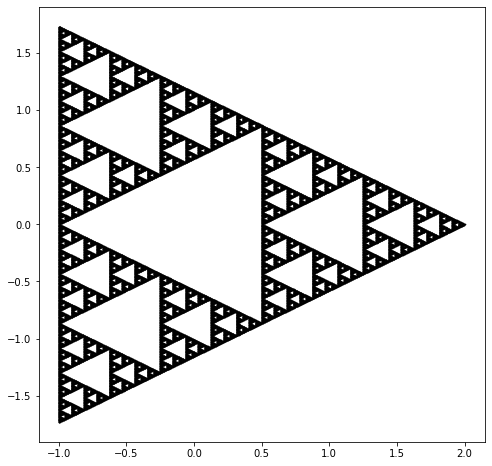}
\caption{Based upon \cite{Chistyakov}, we construct an embedding $\mathfrak{f}:\mathbb{Z}_p\rightarrow \mathbb{R}^2$. The figure shows the images of $\mathfrak{f}(\mathbb{Z}_2)$ and $\mathfrak{f}(\mathbb{Z}_3)$. This computation requires a truncation of the $p$-adic integers. We use $\mathbb{Z}_2/2^{14}\mathbb{Z}_2$ and $\mathbb{Z}_3/3^{10}\mathbb{Z}_3$, respectively.}

%\caption{ The figure contains the visualizations of the fractals
%$\mathbb{Z}_{2}$ (approximated as $\mathbb{Z}_{2}/2^{12}\mathbb{Z}_{2}$) and
%$\mathbb{Z}_{3}$ (approximated as $\mathbb{Z}_{3}/3^{10}\mathbb{Z}_{3}$).
%These fractals were constructed using the results of \cite{Chistyakov}.}%
\label{Figure 2}%
\end{center}
\end{figure}

The unit ball $\mathbb{Z}_{p}$ is an infinite rooted tree with fractal
structure; see  \Cref{Figure 2}. Section \ref{Section6} (Appendix A) provides review of the basic aspects of the $p$-adic analysis required
here. We note that the word `field' will be used here in two different
contexts throughout the article. In a mathematical context, we refer to
algebraic fields; in a physical context, we refer to Euclidean quantum fields.

\section{\label{Section2}Non-Archimedean $\left\{  \boldsymbol{v}%
,\boldsymbol{h}\right\}  ^{4}$-statistical field theories}

We fix $a\left(  x\right)  $, $b(x)$, $c(x)$, $d(x)\in\mathcal{D}%
(\mathbb{Z}_{p})$, $e\in\mathbb{R}$, and an integrable function $w\left(
x\right)  :\mathbb{Z}_{p}\rightarrow\mathbb{R}$ . A $p$\textit{-adic
continuous Boltzmann machine (or a }$p$\textit{-adic continuous
BM)} is a statistical field theory involving two scalars fields $\boldsymbol{v}$, $\boldsymbol{h}$. The function $\boldsymbol{v}%
(x)\in\mathcal{D}(\mathbb{Z}_{p})$ is called the \textit{visible field} and
the function $\boldsymbol{h}(x)\in\mathcal{D}(\mathbb{Z}_{p})$ is called the
\textit{hidden field}. We assume that the field $\left\{  \boldsymbol{v},\boldsymbol{h}%
\right\}  $ performs thermal fluctuations and that the expectation value
of the field is zero.

%EndExpansion
The size of the fluctuations is controlled by an energy functional of the form%
\[
E(\boldsymbol{v},\boldsymbol{h};\boldsymbol{\theta}):=E(\boldsymbol{v}%
,\boldsymbol{h})=E_{0}\left(  \boldsymbol{v},\boldsymbol{h}\right)
+E_{\text{int}}\left(  \boldsymbol{v},\boldsymbol{h}\right)  ,
\]
where $\boldsymbol{\theta}=\left(  w,a,b,c,d,e\right)  $. The first term%
\[
E_{0}\left(  \boldsymbol{v},\boldsymbol{h}\right)  =-%
%TCIMACRO{\dint \limits_{\mathbb{Z}_{p}}}%
%BeginExpansion
{\displaystyle\int\limits_{\mathbb{Z}_{p}}}
%EndExpansion
a(x)\boldsymbol{v}\left(  x\right)  dx-%
%TCIMACRO{\dint \limits_{\mathbb{Z}_{p}}}%
%BeginExpansion
{\displaystyle\int\limits_{\mathbb{Z}_{p}}}
%EndExpansion
b(x)\boldsymbol{h}\left(  x\right)  dx+\frac{e}{2}%
%TCIMACRO{\dint \limits_{\mathbb{Z}_{p}}}%
%BeginExpansion
{\displaystyle\int\limits_{\mathbb{Z}_{p}}}
%EndExpansion
\boldsymbol{v}^{2}\left(  x\right)  dx+\frac{e}{2}%
%TCIMACRO{\dint \limits_{\mathbb{Z}_{p}}}%
%BeginExpansion
{\displaystyle\int\limits_{\mathbb{Z}_{p}}}
%EndExpansion
\boldsymbol{h}^{2}\left(  x\right)  dx
\]
is an analogue of the \textit{free-field energy}. The second term%
\[
E_{\text{int}}\left(  \boldsymbol{v},\boldsymbol{h}\right)  =-%
%TCIMACRO{\diint \limits_{\mathbb{Z}_{p}\times\mathbb{Z}_{p}}}%
%BeginExpansion
{\displaystyle\iint\limits_{\mathbb{Z}_{p}\times\mathbb{Z}_{p}}}
%EndExpansion
\boldsymbol{h}\left(  y\right)  w\left(  x-y\right)  \boldsymbol{v}\left(
x\right)  dxdy+%
%TCIMACRO{\dint \limits_{\mathbb{Z}_{p}}}%
%BeginExpansion
{\displaystyle\int\limits_{\mathbb{Z}_{p}}}
%EndExpansion
c(x)\boldsymbol{v}^{4}\left(  x\right)  dx+%
%TCIMACRO{\dint \limits_{\mathbb{Z}_{p}}}%
%BeginExpansion
{\displaystyle\int\limits_{\mathbb{Z}_{p}}}
%EndExpansion
d(x)\boldsymbol{h}^{4}\left(  x\right)  dx
\]
is an analogue of the \textit{interaction energy}. The results presented in this section are valid for more general functionals in which the first term in $E_{\text{int}}(\boldsymbol{v},\boldsymbol{h})$ is replaced by  
\[%
%{\textstyle\iint\nolimits_{\mathbb{Z}_{p}\times\mathbb{Z}_{p}}}
{\displaystyle\iint\limits_{\mathbb{Z}_{p}\times\mathbb{Z}_{p}}}
\boldsymbol{h}\left(  y\right)  w\left(  x,y\right)  \boldsymbol{v}\left(
x\right)  dxdy.
\]
The motivation behind the definition of the energy functionals
$E(\boldsymbol{v},\boldsymbol{h})$ is that the discretizations of these
functionals give the energy functionals considered in \cite{Batchits et al},
\cite{Le roux et al 1}, \cite{Honglak et al}-\cite{Le roux et al 2}.

All  the thermodynamic  properties of the system are described by the partition
 function of the fluctuating fields, which is defined as
\[
Z^{\text{phys}} (\boldsymbol{\theta})=%
%TCIMACRO{\dint }%
%BeginExpansion
{\displaystyle\int}
%EndExpansion
d\boldsymbol{v}d\boldsymbol{h}\text{ }e^{-\frac{E(\boldsymbol{v}%
,\boldsymbol{h})}{K_{B}T}},
\]
where $K_{B}$ is the Boltzmann constant and $T$ is the temperature constant. We
normalize in such a way that $K_{B}T=1$. The measure $d\boldsymbol{v}%
d\boldsymbol{h}$ is ill-defined. However, it is expected that such measure can be
defined rigorously\ by a limit process. The statistical field theory
corresponding to the energy functional $E(\boldsymbol{v},\boldsymbol{h}%
;\boldsymbol{\theta})$ is defined as the probability measure%
\[
\boldsymbol{P}^{\text{phys}}(\boldsymbol{v},\boldsymbol{h};\boldsymbol{\theta
})=d\boldsymbol{v}d\boldsymbol{h}\frac{\exp\left(  -E(\boldsymbol{v}%
,\boldsymbol{h})\right)  }{Z^{\text{phys}}},
\]
on the space  $\mathcal{D}(\mathbb{Z}_{p})\times\mathcal{D}%
(\mathbb{Z}_{p})$.

The information about the local properties of the system is contained in the
\textit{correlation functions }$G_{\mathbb{I},\mathbb{K}}^{\left(  n\right)
}\left(  x_{1},\ldots,x_{n}\right)  $ of the field $\left\{  \boldsymbol{v}%
,\boldsymbol{h}\right\}  $: for $n\geq1$, and two disjoint subsets
$\mathbb{I}$, $\mathbb{K}\subset\left\{  1,2,\ldots,n\right\}  $, with
$\mathbb{I}%
%TCIMACRO{\tcoprod }%
%BeginExpansion
{\textstyle\coprod}
%EndExpansion
\mathbb{K}=\left\{  1,2,\ldots,n\right\}  $, where ${\textstyle\coprod}$ is the disjoint union,  we set
\begin{gather*}
G_{\mathbb{I},\mathbb{K}}^{\left(  n\right)  }\left(  x_{1},\ldots
,x_{n}\right)  =\left\langle
%TCIMACRO{\dprod \limits_{i\in\mathbb{I}}}%
%BeginExpansion
{\displaystyle\prod\limits_{i\in\mathbb{I}}}
%EndExpansion
\boldsymbol{v}\left(  x_{i}\right)  \text{ }%
%TCIMACRO{\dprod \limits_{j\in\mathbb{K}}}%
%BeginExpansion
{\displaystyle\prod\limits_{j\in\mathbb{K}}}
%EndExpansion
\boldsymbol{h}\left(  x_{j}\right)  \right\rangle \\
:=\frac{1}{Z^{\text{phys}}}%
%TCIMACRO{\dint }%
%BeginExpansion
{\displaystyle\int}
%EndExpansion
d\boldsymbol{v}d\boldsymbol{h}\text{ }%
%TCIMACRO{\dprod \limits_{i\in\mathbb{I}}}%
%BeginExpansion
{\displaystyle\prod\limits_{i\in\mathbb{I}}}
%EndExpansion
\boldsymbol{v}\left(  x_{i}\right)  \text{ }%
%TCIMACRO{\dprod \limits_{j\in\mathbb{K}}}%
%BeginExpansion
{\displaystyle\prod\limits_{j\in\mathbb{K}}}
%EndExpansion
\boldsymbol{h}\left(  x_{j}\right)  \text{ }e^{-E(\boldsymbol{v}%
,\boldsymbol{h})}.
\end{gather*}
These functions are also called the $n$-point Green functions. 

To study these functions, one introduces two auxiliary external fields $J_{0}(x),$
$J_{1}(x)\in\mathcal{D}(\mathbb{Z}_{p})$ called \textit{currents, }and adds to
the energy functional $E$ as a linear interaction energy of these currents
with the field $\left\{  \boldsymbol{v},\boldsymbol{h}\right\}  $,%
\[
E_{\text{source}}(\boldsymbol{v},\boldsymbol{h},J_{0},J_{1})=-%
%TCIMACRO{\dint \limits_{\mathbb{Z}_{p}}}%
%BeginExpansion
{\displaystyle\int\limits_{\mathbb{Z}_{p}}}
%EndExpansion
J_{0}(x)\boldsymbol{v}\left(  x\right)  dx-%
%TCIMACRO{\dint \limits_{\mathbb{Z}_{p}}}%
%BeginExpansion
{\displaystyle\int\limits_{\mathbb{Z}_{p}}}
%EndExpansion
J_{1}(x)\boldsymbol{h}\left(  x\right)  dx,
\]
now the energy functional is $E(\boldsymbol{v},\boldsymbol{h},J_{0}%
,J_{1})=E\left(  \boldsymbol{v},\boldsymbol{h}\right)  +E_{\text{source}%
}(\boldsymbol{v},\boldsymbol{h},J_{0},J_{1})$. The partition function formed
with this energy is%
\[
Z(J_{0},J_{1})=\frac{1}{Z_{0}^{\text{phys}}}%
%TCIMACRO{\dint }%
%BeginExpansion
{\displaystyle\int}
%EndExpansion
d\boldsymbol{v}d\boldsymbol{h}\text{ }e^{-E(\boldsymbol{v},\boldsymbol{h}%
,J_{0},J_{1})},
\]
where
\[
Z_{0}^{\text{phys}}=%
%TCIMACRO{\dint }%
%BeginExpansion
{\displaystyle\int}
%EndExpansion
d\boldsymbol{v}d\boldsymbol{h}\text{ }e^{-E_{0}(\boldsymbol{v},\boldsymbol{h}%
)}.
\]
The functional derivatives of $Z(J_{0},J_{1})$ with respect to $J_{0}(x)$,
$J_{1}(x)$ evaluated at $J_{0}=0$, $J_{1}=0$\ give the correlation functions
of the system:%
\[
G_{\mathbb{I},\mathbb{K}}^{\left(  n\right)  }\left(  x_{1},\ldots
,x_{n}\right)  =\frac{1}{Z}\left[
%TCIMACRO{\tprod \limits_{i\in\mathbb{I}}}%
%BeginExpansion
{\textstyle\prod\limits_{i\in\mathbb{I}}}
%EndExpansion
\frac{\delta}{\delta J_{0}\left(  x_{i}\right)  }\text{ }%
%TCIMACRO{\tprod \limits_{j\in\mathbb{K}}}%
%BeginExpansion
{\textstyle\prod\limits_{j\in\mathbb{K}}}
%EndExpansion
\frac{\delta}{\delta J_{1}\left(  x_{j}\right)  }Z(J_{0},J_{1})\right]
_{\substack{J_{0}=0\\J_{1}=0}},
\]
where $Z=\frac{Z^{\text{phys}}}{Z_{0}^{\text{phys}}}$. The functional
$Z(J_{0},J_{1})$ is called the \textit{generating functional} of the theory.

The description of the $\left\{  \boldsymbol{v},\boldsymbol{h}\right\}  ^{4}%
$-SFTs presented above is based in the classical version of these theories
\cite{Kleinert et al}-\cite{Mussardo}. In \cite{Zuniga-RMP-2022}, a
mathematically rigorous formulation of $\phi^{4}$-SFTs is presented, the
fields are functions from $\mathbb{Q}_{p}^{N}$ into $\mathbb{R}$, with $N$
arbitrary. We expect that this theory can be extended to the $\left\{
\boldsymbol{v},\boldsymbol{h}\right\}  ^{4}$-SFTs presented here.

\section{\label{Section3}Discrete SFTs and $p$-adic discrete Boltzmann
machines}

A central
difference between the $p$-adic STFs and the classical ones is that in the
$p$-adic case, the discretization process can be carried out in an easy
rigorous way. More specifically, the discretization of a $p$-adic SFT is constructed by restricting the energy
functional $E(\boldsymbol{v},\boldsymbol{h};\boldsymbol{\theta})$ to a finite
dimensional vector subspace $\mathcal{D}^{l}(\mathbb{Z}_{p})$ of the space of
test functions $\mathcal{D}(\mathbb{Z}_{p})$. 
The test functions in $\mathcal{D}^{l}(\mathbb{Z}_{p})$ have the
form 
\[
\varphi\left(  x\right)  =%
%TCIMACRO{\tsum \limits_{i\in G_{l}}}%
%BeginExpansion
{\textstyle\sum\limits_{i\in G_{l}}}
%EndExpansion
\varphi\left(  i\right)  \Omega\left(  p^{l}\left\vert x-i\right\vert
_{p}\right), \quad
  \varphi\left(  i\right)  \in\mathbb{R}, \]
 where
$i\boldsymbol{=}i_{0}+i_{1}p+\ldots+i_{l-1}p^{l-1}\in G_{l}=\mathbb{Z}%
_{p}/p^{l}\mathbb{Z}_{p}$, $l\geq1$, and $\Omega\left(  p^{l}\left\vert
x-i\right\vert _{p}\right)  $ is the characteristic function of the ball
$B_{-l}(i)$. Here, it is important to notice that $G_{l}$ is a
finite, Abelian, additive group.
In the $p$-adic world, the
discrete functions are a particular case of the $p$-adic continuous functions,
more precisely, $\mathcal{D}(\mathbb{Z}_{p})=\cup_{l \in \mathbb{N}}\mathcal{D}%
^{l}(\mathbb{Z}_{p})$ and $\mathcal{D}^{l}(\mathbb{Z}_{p})\subset$
$\mathcal{D}^{l+1}(\mathbb{Z}_{p})$. There is no Archimedean counterpart of
this result. 

By taking $\boldsymbol{v},\boldsymbol{h}\in\mathcal{D}^{l}(\mathbb{Z}_{p})$
and $l$ sufficiently large, the restriction $E_{l}\left(  \boldsymbol{v}%
,\boldsymbol{h};\boldsymbol{\theta}\right)  $ of the energy functional
$E(\boldsymbol{v},\boldsymbol{h};\boldsymbol{\theta})$ to $\mathcal{D}%
^{l}(\mathbb{Z}_{p})$ has the form
\begin{gather}\label{energy-discretization}
E_{l}\left(  \boldsymbol{v},\boldsymbol{h};\boldsymbol{\theta}\right)  =-%
%TCIMACRO{\dsum \limits_{j\in G_{l}}}%
%BeginExpansion
{\displaystyle\sum\limits_{j\in G_{l}}}
%EndExpansion%
%TCIMACRO{\dsum \limits_{k\in G_{l}}}%
%BeginExpansion
{\displaystyle\sum\limits_{k\in G_{l}}}
%EndExpansion
w_{k}v_{j+k}h_{j}-%
%TCIMACRO{\dsum \limits_{j\in G_{l}}}%
%BeginExpansion
{\displaystyle\sum\limits_{j\in G_{l}}}
%EndExpansion
a_{j}v_{j}-%
%TCIMACRO{\dsum \limits_{j\in G_{l}}}%
%BeginExpansion
{\displaystyle\sum\limits_{j\in G_{l}}}
%EndExpansion
b_{j}h_{j}+\frac{e}{2}%
%TCIMACRO{\dsum \limits_{j\in G_{l}^{N}}}%
%BeginExpansion
{\displaystyle\sum\limits_{j\in G_{l}^{N}}}
%EndExpansion
v_{j}^{2}+\frac{e}{2}%
%TCIMACRO{\dsum \limits_{j\in G_{l}}}%
%BeginExpansion
{\displaystyle\sum\limits_{j\in G_{l}}}
%EndExpansion
h_{j}^{2}\\ \nonumber
+%
%TCIMACRO{\dsum \limits_{j\in G_{l}}}%
%BeginExpansion
{\displaystyle\sum\limits_{j\in G_{l}}}
%EndExpansion
c_{j}v_{j}^{4}+%
%TCIMACRO{\dsum \limits_{j\in G_{l}}}%
%BeginExpansion
{\displaystyle\sum\limits_{j\in G_{l}}}
%EndExpansion
d_{j}h_{j}^{4}\text{.}%
\end{gather}
We refer \Cref{Section7} (Appendix B) for further details of this calculation. From now on, we refer \cref{energy-discretization} as the energy functional for a discrete\ $\left\{
\boldsymbol{v},\boldsymbol{h}\right\}  ^{4}$-STF.

In the general case, $E_{l}\left(
\boldsymbol{v},\boldsymbol{h};\boldsymbol{\theta}\right)  $ is a $p$-adic
analogue of the $\left\{  \boldsymbol{v},\boldsymbol{h}\right\}  ^{4}$ neural
networks introduced in \cite{Batchits et al}. In this article,  
each hidden state $h_{j}$ and visible state $v_{i}$ are interacted through a weight $w_{ij}$. Therefore, requires the number of $G_{l}^{2}$  for weights $w$, whereas our counterpart requires only $G_{l}$.   See  Section \ref{Section5}, for further discussion.

We now attach to the discrete energy functional a
$p$-adic discrete BM.
For any visible and hidden states, $\boldsymbol{v}=\left[  v_{i}\right]  _{i\in G_{l}}$ and
\textit{\ }$\boldsymbol{h}=\left[  h_{i}\right]  _{i\in G_{l}}$, the Boltzmann
probability distribution attached to $E_{l}\left(  \boldsymbol{v}%
,\boldsymbol{h};\boldsymbol{\theta}\right)  $ is given by%
\[
\boldsymbol{P}_{l}(\boldsymbol{v},\boldsymbol{h};\boldsymbol{\theta}%
)=\frac{\exp\left(  -E_{l}\left(  \boldsymbol{v},\boldsymbol{h}%
;\boldsymbol{\theta}\right)  \right)  }{{\sum\limits_{\boldsymbol{v}%
,\boldsymbol{h}}}\exp\left(  -E_{l}\left(  \boldsymbol{v},\boldsymbol{h}%
;\boldsymbol{\theta}\right)  \right)  }.
\]
When there is no risk of confusion, we will omit $\boldsymbol{\theta}$ in the
notations. When $c_{j}=d_{j}=0$ for all $j\in G_{l}$, the energy functional
$E_{l}\left(  \boldsymbol{v},\boldsymbol{h};\boldsymbol{\theta}\right)  $
corresponds to a $p-$adic analogue of the convolutional deep belief networks
studied in \cite{Honglak et al}. 

We note that the energy functional $E_{l}$ has translational symmetry,
i.e., $E_{l}$ is invariant under the transformations $j\rightarrow j+j_{0}$,
$k\rightarrow k+k_{0}$, for any $j_{0}$, $k_{0}\in G_{l}$. This transformation is well-defined since $G_{l}$ is an
additive group.
 In the case of
applications to image processing, 
the group property also implies that the convolution operation does not alter image dimensions. 
The convolutional $p$-adic discrete BMs
introduced here are a specific type of deep Boltzmann machines (DBMs) (also called deep belief
networks DBNs).

\section{\label{Section4}Experimental Results}

We implement a $p$-adic discrete Boltzmann machine for processing binary
images, then $\boldsymbol{v},\boldsymbol{h}:\mathbb{Z}_{p}\rightarrow\left\{
0,1\right\}  \subset\mathbb{R}$, in this case, we use the following energy functional:
\[
E_{l}\left(  \boldsymbol{v},\boldsymbol{h};\theta\right)  =-{\sum\limits_{j\in
G_{l}}}{\sum\limits_{\substack{k\in G_{l}\\|k|_{p}\leq p^{-N}}}}w_{k}%
v_{j+k}h_{j}-{\sum\limits_{j\in G_{l}}}a_{j}v_{j}-{\sum\limits_{j\in G_{l}}%
}b_{j}h_{j}\text{,}%
\]
for some natural number $0\leq N \leq l$. Note that, comparing to \Cref{energy-discretization}, the quadratic and biquadratic terms are omitted since they do not play any role in
the case in which $\boldsymbol{v},\boldsymbol{h}$ are binary variables. The
condition $N \le l$ implies that the convolution operation is restricted to 
 a small neighborhood of radius $p^{-N}$ for each pixel. The condition $N=0$ means that the
convolution involves all the points in the image.

Our numerical experiment is based on the MNIST dataset, where each image is considered as a sample of the visible state. Our first task is to train the network to maximize the log-likelihood of the visible states. We choose $p=3$ and $l=6$ since image dimension is
$3^{3}\times3^{3}$. In general, $p$ and $l$ depend on the size of the
images to be processed. Typically $p$ is chosen as a small prime number, e.g., $2$ or $3$.

\begin{figure}[h]
\begin{center}
\includegraphics[
width=0.5\textwidth
]%
{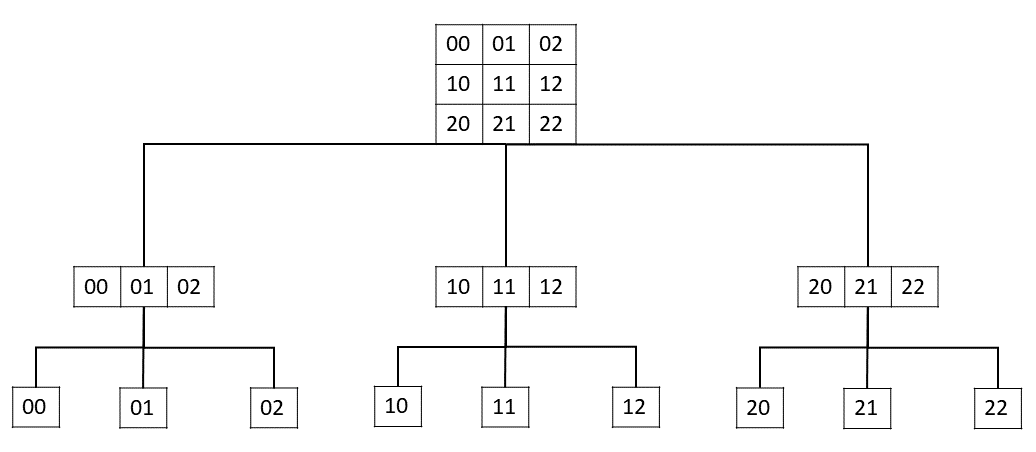}%
\caption{The construction of the tree corresponding to a $3\times3$ image.}%
\label{Figure 3}%
\end{center}
\end{figure}

\begin{figure}[h]
\begin{center}
\includegraphics[
width=0.45\textwidth
]%
{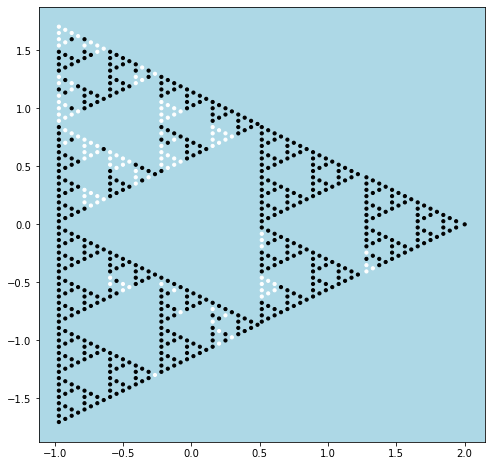}%
\includegraphics[
width=0.43\textwidth
]%
{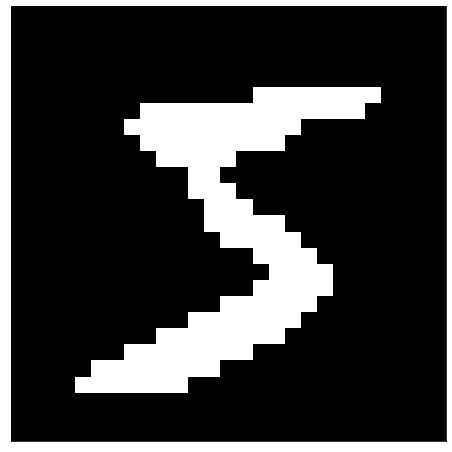}

\caption{The processing of date using $p$-adic networks requires the transformation of the actual data to its $p$-adic form. The left image shows a visualization of the test function corresponding to right image. The visualization uses the embedding $\mathfrak{f}:\mathbb{Z}_p\rightarrow \mathbb{R}^2$ showed in
Figure \ref{Figure 2}.}
%\caption{The $p$-adic networks process require the transformation of actual data to its $p$-adic form. The left image shows a visualization of the test function corresponding to right image. The visualization uses the embedding $\mathfrak{f}:\mathbb{Z}_p\rightarrow \mathbb{R}^2$ introduced in \cite{Chistyakov}.}%
\label{Figure 4}%
\end{center}
\end{figure}

To tune the parameters $a_{j}$, $b_{j}$, $w_{j}$,
$j\in G_{6}$ of the network, we first transform each image $I$
into a test function $\text{Test}\left(  I\right)  \in\mathcal{D}^{6}(\mathbb{Z}%
_{3})$.
The test function $\text{Test}(I)$ is defined in terms of the tree structure of
$G_{6}$ in the following way: we define $I$ as the root of the tree. Later we
divide $I$ into three horizontal even slices (sub-images). These sub-images are the
vertices level $1$, and they are the children of $I$. Each sub-image $I_{j}$ at
level $1$ is then divided vertically into $3$ sub-images; these are the children of
the $I_{j}$. All the $3^{2}$ new sub-images correspond to the vertices at
level $2$. We repeat this process until reaching level  $6$. At level $6$, each
vertex corresponds to a pixel, and we denote by value $I_{i}$ for $i\in G_{6}$. Then $\text{Test}(I)$
is defined as $\sum_{i\in G_{6}}I_{i}\Omega(3^{6}|x-i|_{3})$. See \Cref{Figure 3} for the construction
of the tree corresponding to a $3^{2}\times3^{2}$ image. Figure \ref{Figure 4} shows the graph of a test function corresponding to an image from the MNIST data set. For further details, the reader may consult the Appendix in \cite{ZZ2}.

%%%%%% Figure 3

%%%%%% Figure 4

\begin{figure}[h]
\begin{center}
\includegraphics[
width=0.5\textwidth
]%
{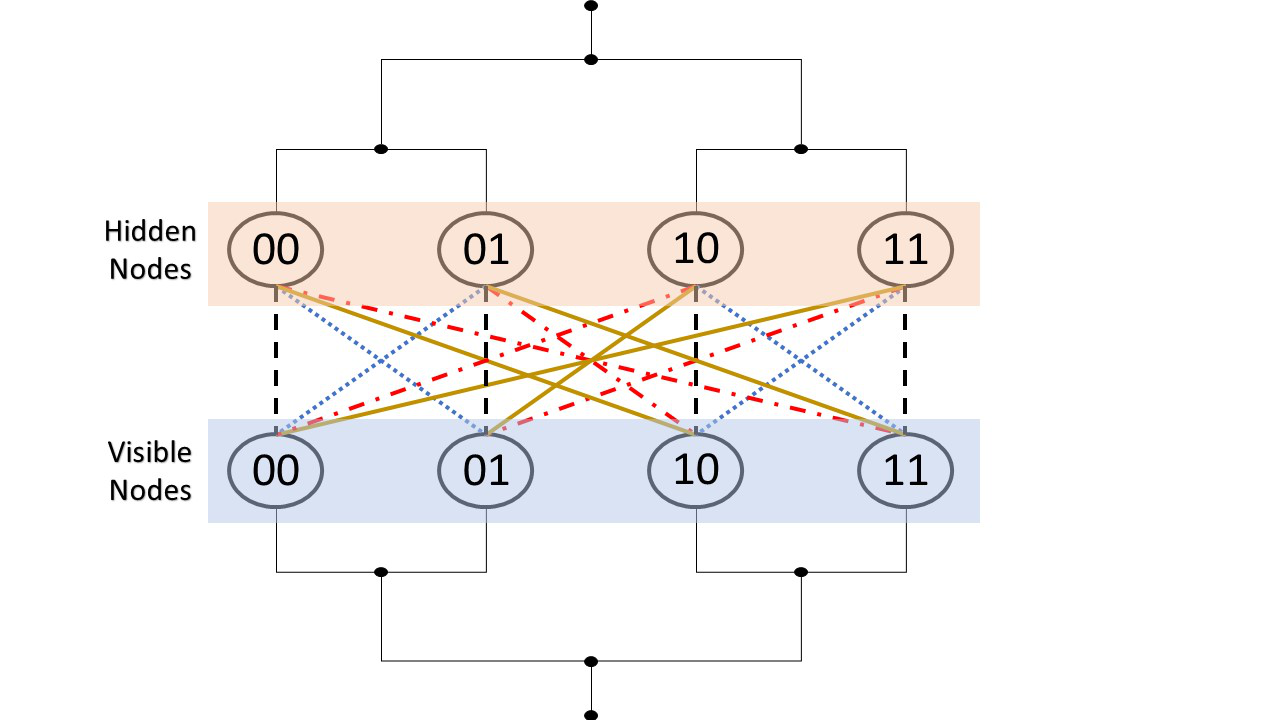}%
\caption{A p-adic RBM for $p=2, l=2$ }%
\label{Figure 5}%
\end{center}
\end{figure}

We note that in a $p$-adic discrete deep RBM, the visible and hidden states are
functions on a finite tree. Only the vertices at the top level, which are marked as orange and
blue balls are allowed to have states. See \Cref{Figure 5} for the case $p=2, l=2$.  The remaining trees' vertices (see the black dots in Figure \ref{Figure 5}) only codify the hierarchical relation between states. The visible and hidden vertices connected by the same type of lines share the same weight $w$.

\begin{figure}[h]
\begin{center}
\subfloat[]{\includegraphics[
width=0.4\textwidth
]{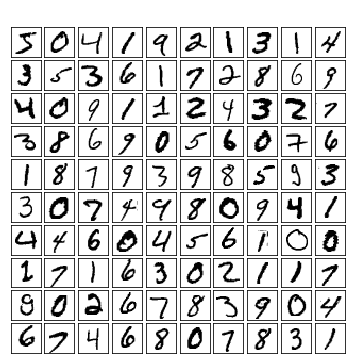}
}
\subfloat[]{\includegraphics[
width=0.45\textwidth
]{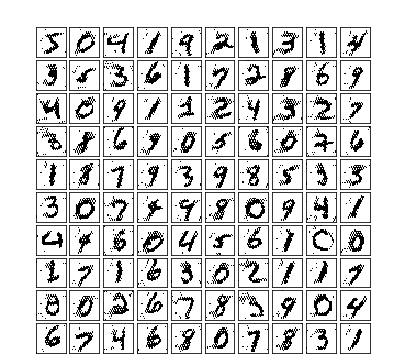}}
\newline
\subfloat[]{\includegraphics[
width=0.45\textwidth
]{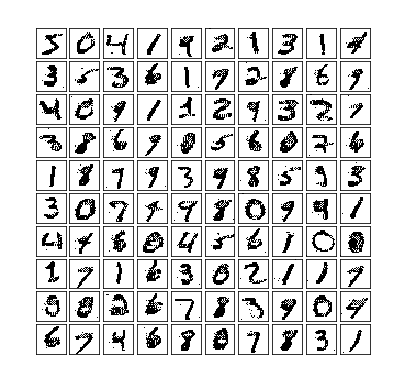}}
\subfloat[]{\includegraphics[
width=0.45\textwidth
]{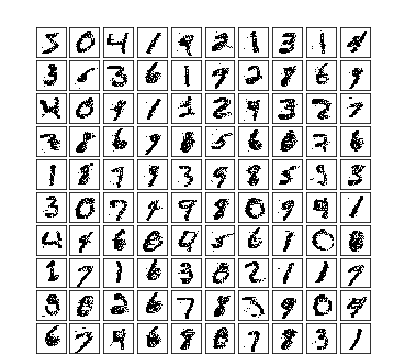}}
\caption{(a) is the original input image. (b) is the reconstructed image  using \ Gibbs
sampling with \ $w$ \ having $\mathbb{Z}_{3}$ as support. (c) is the reconstructed
image with $w$ supported in $3^{3}\mathbb{Z}_{3}$. (d) is the reconstructed
image with $w$ supported in $3^{4}\mathbb{Z}_{3}$.}%
\label{Figure 6}%
\end{center}
\end{figure}

We adapted the contrastive divergence learning method to the $p$-adic framework,
\cite{Fischer-Igel}. The technical details are presented in Section
\ref{Sec:CD_learning}. We implement two different types of networks. In the first
type, the function $w_{k}$ is supported in the entire tree $G_{6}$; in the
second type, the function $w_{k}$ is supported in a proper subset of the tree
$G_{6}$. We use the full MNIST hadwritten digits, without considering labels,
to train a six layer $3$-adic feature detector. The results are show in Figure
\ref{Figure 6}.

%%%%%%%%%%% Figure5

After the processing by the network ends, it is necessary to
transform the test function into an image.

\section{\label{Section5}Conclusions}

The standard RBMs \cite{Fischer-Igel} and the $\phi^{4}$- neural networks
introduced in \cite{Batchits et al}\ are particular classes of $p$-adic
discrete BMs. Indeed, if $w(x,y)$ is a test function and the interaction
between the visible and hidden field has the form $-%
%TCIMACRO{\tiint \nolimits_{\mathbb{Z}_{p}\times\mathbb{Z}_{p}}}%
%BeginExpansion
{\textstyle\iint\nolimits_{\mathbb{Z}_{p}\times\mathbb{Z}_{p}}}
%EndExpansion
\boldsymbol{h}\left(  y\right)  w\left(  x,y\right)  \boldsymbol{v}\left(
x\right)  dxdy$, then in the corresponding discrete energy functional, after a
suitable rescaling of the weights $w_{k,j}$, the interaction of the visible
and hidden states takes the form $-\sum_{j\in G_{l}}\sum_{k\in G_{l}}%
w_{k,j}v_{k}h_{j}$.  Here $\left[  w_{k,j}\right]  $ is an ordinary matrix,
which means that its entries $w_{k,j}$ do not depend on the algebraic
structure of $G_{l}$ neither on its topology.

In the case in which the interaction between the visible and hidden field has
the form $-%
%TCIMACRO{\tiint \nolimits_{\mathbb{Z}_{p}\times\mathbb{Z}_{p}}}%
%BeginExpansion
{\textstyle\iint\nolimits_{\mathbb{Z}_{p}\times\mathbb{Z}_{p}}}
%EndExpansion
\boldsymbol{h}\left(  y\right)  w\left(  x-y\right)  \boldsymbol{v}\left(
x\right)  dxdy$, then corresponding discrete energy functional depends on the
group structure of $G_{l}$, and the corresponding neural network is a
particular case of a DBM, see Figure \ref{Figure 5}.

The condition $l\geq l_{0}$, for some constant $l_{0}$, means that a $p$-adic
discrete BM\ admits copies arbitratly large, this is a consequence of the fact
of $p$-adic numbers has a tree-like structure. We expect that for $l$ sufficiently large
the statistical properties of the network can be studied using a $p$-adic
continuous SFT. 

Our numerical experiments show that $p$-adic discrete convolutional deep BMs
alone can be used to process real data, this opens the possibility of using
these networks as layers in specialized NNs.

In \cite{Zuniga1} the first author conjectured that the limit%
\begin{equation}
\frac{e^{-E(\boldsymbol{v},\boldsymbol{h})}}{Z^{\text{phys}}}d\boldsymbol{v}%
d\boldsymbol{h}=\lim_{l\rightarrow\infty}\boldsymbol{P}_{l}(\boldsymbol{v}%
,\boldsymbol{h})\text{ }d^{\#G_{l}}\boldsymbol{v}\text{ }d^{\#G_{l}%
}\boldsymbol{h} \label{limit}%
\end{equation}
exists in some sense, here $d^{\#G_{l}}\boldsymbol{x}$ denotes the Lebesgue
measure of $\mathbb{R}^{\#G_{l}}$. Then the correlation between the network
activity in different regions of the underlaying tree $G_{l}$ can be
understood by computing the correlations functions of the corresponding
continuous SFT.

For practical applications the NNs should be discrete entities. This type of
NNs naturally correspond with discrete SFTs. To use Euclidean QFT to study
NNs, it is convenient to have continuous versions of these networks. Thus, a
clear way of passing between discrete STFs to continuous ones is required.
The existence of the limit \ref{limit} is a very difficult problem in
classical QFT. In \cite{Zuniga-RMP-2022} the existence of this limit was
established for $p$-adic $\phi^{4}$- theories involving one scalar field, we
expect that these techniques can be extended to the case QFTs considered here.

It is widely accepted in the artificial intelligence community that the
probability distributions $\boldsymbol{P}_{l}(\boldsymbol{v},\boldsymbol{h})$
should approximate any finite probability distribution very well, which means
that the corresponding NNs are universal approximators. We argue that this
property is connected with the topology and the structure of the NNs, and that
the problem of designing good architectures for NNs is out of the scope of
QFT. We expect that QFT techniques will be useful to understand the
qualitative behavior of large NNs, which can be well-approximated as
`continuos' NNs.

The study of the correspondence between $p$-adic Euclidean QFTs and NNs is
just starting. We envision that the next step is to develop perturbative
calculations of the correlation functions via Feynman diagrams, to study
connections with Ginzburg--Landau theory, and to develop practical
applications of the $p$-adic convolutional BMs.%

%%%%%%%%%% Figure 6
%EndExpansion

\section{\label{Section6}Apendix A: Basic facts on $p$-adic analysis}

In this section we review some basic results on $p$-adic analysis required in
this article. For a detailed exposition on $p$-adic analysis the reader may
consult \cite{V-V-Z}, \cite{A-K-S}-\cite{Taibleson}. For a quick review of
$p$-adic analysis the reader may consult \cite{Bocardo-Zuniga-2}.

\subsection{The field of $p$-adic numbers}

The field of $p-$adic numbers $\mathbb{Q}_{p}$ is defined as the completion of
the field of rational numbers $\mathbb{Q}$ with respect to the $p-$adic norm
$|\cdot|_{p}$, which is defined as
\begin{equation}\label{p-norm}
|x|_{p}=%
\begin{cases}
0 & \text{if }x=0\\
p^{-\gamma} & \text{if }x=p^{\gamma}\dfrac{a}{b},
\end{cases}
\end{equation}
where $a$ and $b$ are integers coprime with $p$. The integer $\gamma
=ord_{p}(x):=ord(x)$, with $ord(0):=+\infty$, is called the\textit{\ }%
$p-$\textit{adic order of} $x$. The metric space $\left(  \mathbb{Q}%
_{p},\left\vert \cdot\right\vert _{p}\right)  $ is a complete ultrametric
space. Ultrametric means that $\left\vert x+y\right\vert _{p}\leq\max\left\{
\left\vert x\right\vert _{p},\left\vert y\right\vert _{p}\right\}  $. As a
topological space $\mathbb{Q}_{p}$\ is homeomorphic to a Cantor-like subset of
the real line, see, e.g., \cite{V-V-Z}, \cite{Chistyakov}, \cite{A-K-S}.

Any $p-$adic number $x\neq0$ has a unique expansion of the form
\[
x=p^{ord(x)}\sum_{j=0}^{\infty}x_{j}p^{j},
\]
where $x_{j}\in\{0,1,2,\dots,p-1\}$ and $x_{0}\neq0$. In addition, for any
$x\in\mathbb{Q}_{p}\smallsetminus\left\{  0\right\}  $ 
%can be represented
%uniquely as
we have
\[
\left\vert x\right\vert
_{p}=p^{-ord(x)}.
\]

\subsection{Topology of $\mathbb{Q}_{p}$}

For $r\in\mathbb{Z}$, denote by $B_{r}(a)=\{x\in\mathbb{Q}_{p};\left\vert
x-a\right\vert _{p}\leq p^{r}\}$ \textit{the ball of radius }$p^{r}$
\textit{with center at} $a\in\mathbb{Q}_{p}$, and take $B_{r}(0):=B_{r}$. {The
ball $B_{0}$ equals \textit{the ring of }$p-$\textit{adic integers
}$\mathbb{Z}_{p} $.} The balls are both open and closed subsets in
$\mathbb{Q}_{p}$. We use $\Omega\left(  p^{-r}\left\vert x-a\right\vert
_{p}\right)  $ to denote the characteristic function of the ball $B_{r}(a)$.
Two balls in $\mathbb{Q}_{p}$ are either disjoint or one is contained in the
other. As a topological space $\left(  \mathbb{Q}_{p},\left\vert
\cdot\right\vert _{p}\right)  $ is totally disconnected, i.e., the only
connected \ subsets of $\mathbb{Q}_{p}$ are the empty set and the points. A
subset of $\mathbb{Q}_{p}$ is compact if and only if it is closed and bounded
in $\mathbb{Q}_{p}$, see e.g. \cite[Section 1.3]{V-V-Z}, or \cite[Section
1.8]{A-K-S}. The balls and spheres are compact subsets. Thus $\left(
\mathbb{Q}_{p},\left\vert \cdot\right\vert _{p}\right)  $ is a locally compact
topological space.

\subsubsection{Tree-like structures}

Any $p$-adic integer $i$ admits an expansion of the form $i=i_{k}p^{k}%
+i_{k+1}p^{k+1}+\ldots$ for some $k\geq0, i_{k}\neq0$. The set of $p$-adic
truncated integers modulo $p^{l}$, $l\geq1$, consists of all the integers of
the form $i\boldsymbol{=}i_{0}+i_{1}p+\ldots+i_{l-1}p^{l-1}$. These numbers
form a complete set of representatives for the elements of the additive group
$G_{l}=\mathbb{Z}_{p}/p^{l}\mathbb{Z}_{p}$, which is isomorphic to the set of
integers $\mathbb{Z}/p^{l}\mathbb{Z}$ (written in base $p$) modulo $p^{l}$. By
restricting $\left\vert \cdot\right\vert _{p}$ to $G_{l}$, it becomes a normed
space, and $\left\vert G_{l}\right\vert _{p}=\left\{  0,p^{-\left(
l-1\right)  },\cdots,p^{-1},1\right\}  $. With the metric induced by
$\left\vert \cdot\right\vert _{p}$, $G_{l}$ becomes a finite ultrametric
space. In addition, $G_{l}$ can be identified with the set of branches
(vertices at the top level) of a rooted tree with $l+1$ levels and $p^{l}$
branches. By definition the root of the tree is the only vertex at level $0$.
There are exactly $p$ vertices at level $1$, which correspond with the
possible values of the digit $i_{0}$ in the $p$-adic expansion of $i$. Each of
these vertices is connected to the root by a non-directed edge. At level $k$,
with $2\leq k\leq l+1$, there are exactly $p^{k}$ vertices, \ each vertex
corresponds to a truncated expansion of $i$ of the form $i_{0}+\cdots
+i_{k-1}p^{k-1}$. The vertex corresponding to $i_{0}+\cdots+i_{k-1}p^{k-1}$ is
connected to a vertex $i_{0}^{\prime}+\cdots+i_{k-2}^{\prime}p^{k-2}$ at the
level $k-1$ if and only if $\left(  i_{0}+\cdots+i_{k-1}p^{k-1}\right)
-\left(  i_{0}^{\prime}+\cdots+i_{k-2}^{\prime}p^{k-2}\right)  $ is divisible
by $p^{k-1}$. See Figure \ref{Figure 1}. The balls $B_{-r}(a)=a+p^{r}%
\mathbb{Z}_{p}$ are\ an infinite rooted trees.

\subsection{The Haar measure}\label{Sec: Haar measure}

Since $(\mathbb{Z}_{p},+)$ is a locally compact topological group, there
exists a Haar measure $dx$, which is invariant under translations, i.e.,
$d(x+a)=dx$, \cite{Halmos}. If we normalize this measure by the condition
$\int_{\mathbb{Z}_{p}}dx=1$, then $dx$ is unique. It follows immediately that
\[%
%TCIMACRO{\tint \limits_{B_{r}(a)}}%
%BeginExpansion
{\textstyle\int\limits_{B_{r}(a)}}
%EndExpansion
dx=%
%TCIMACRO{\tint \limits_{a+p^{-r}\mathbb{Z}_{p}}}%
%BeginExpansion
{\textstyle\int\limits_{a+p^{-r}\mathbb{Z}_{p}}}
%EndExpansion
dx=p^{r}%
%TCIMACRO{\tint \limits_{\mathbb{Z}_{p}}}%
%BeginExpansion
{\textstyle\int\limits_{\mathbb{Z}_{p}}}
%EndExpansion
dy=p^{r}\text{, }r\in\mathbb{Z}\text{.}%
\]
In a few ocassions we use the two-dimensional Haar measure $dxdy$ of the
additive group $(\mathbb{Z}_{p}\times\mathbb{Z}_{p},+)$ normalize this measure
by the condition $\int_{\mathbb{Z}_{p}}\int_{\mathbb{Z}_{p}}dxdy=1$. For a
quick review of the integration in the $p$-adic framework the reader may
consult \cite{Bocardo-Zuniga-2} and the references therein.

\subsection{The Bruhat-Schwartz space in the unit ball}

A real-valued function $\varphi$ defined on $\mathbb{Z}_{p}$ is \textit{called
Bruhat-Schwartz function (or a test function)} if for any $x\in\mathbb{Z}_{p}$
there exist an integer $l(x)\in\mathbb{Z}$ such that%
\begin{equation}
\varphi(x+x^{\prime})=\varphi(x)\text{ for any }x^{\prime}\in B_{l(x)}.
\label{local_constancy}%
\end{equation}
The $\mathbb{R}$-vector space of Bruhat-Schwartz functions supported in the
unit ball is denoted by $\mathcal{D}(\mathbb{Z}_{p})$. For $\varphi
\in\mathcal{D}(\mathbb{Z}_{p})$, the largest number $l=l(\varphi)$ satisfying
(\ref{local_constancy}) is called \textit{the exponent of local constancy (or
the parameter of constancy) of} $\varphi$. A function $\varphi$ in
$\mathcal{D}(\mathbb{Z}_{p})$ can be written as%
\[
\varphi\left(  x\right)  =%
%TCIMACRO{\dsum \limits_{j=1}^{M}}%
%BeginExpansion
{\displaystyle\sum\limits_{j=1}^{M}}
%EndExpansion
\varphi\left(  \widetilde{x}_{j}\right)  \Omega\left(  p^{r_{j}}\left\vert
x-\widetilde{x}_{j}\right\vert _{p}\right)  ,
\]
where the $\widetilde{x}_{j}$, $j=1,\ldots,M$, are points in $\mathbb{Z}_{p}$,
the $r_{j}$, $j=1,\ldots,M$, are integers, and $\Omega\left(  p^{r_{j}%
}\left\vert x-\widetilde{x}_{j}\right\vert _{p}\right)  $ denotes the
characteristic function of the ball $B_{-r_{j}}(\widetilde{x}_{j}%
)=\widetilde{x}_{j}+p^{r_{j}}\mathbb{Z}_{p}$.

\section{\label{Section7}Appendix B: Discretization of the energy functional}

The elments of $G_{l}=\mathbb{Z}_{p}/p^{l}\mathbb{Z}_{p}$, $l\geq1$, have the
form $i\boldsymbol{=}i_{0}+i_{1}p+\ldots+i_{l-1}p^{l-1}$, where the $i_{k}%
$s\ are $p$-adic digits. We denote by $\mathcal{D}^{l}(\mathbb{Z}_{p})$ the
$\mathbb{R}$-vector space of all test functions of the form%
\[
\varphi\left(  x\right)  =%
%TCIMACRO{\tsum \limits_{i\in G_{l}}}%
%BeginExpansion
{\textstyle\sum\limits_{i\in G_{l}}}
%EndExpansion
\varphi\left(  i\right)  \Omega\left(  p^{l}\left\vert x-i\right\vert
_{p}\right)  \text{, \ }\varphi\left(  i\right)  \in\mathbb{R}\text{,}%
\]
here $\Omega\left(  p^{l}\left\vert x-i\right\vert _{p}\right)  $ is the
characteristic function of the ball $i+p^{l}\mathbb{Z}_{p}$. Notice that
$\varphi$ is supported on $\mathbb{Z}_{p}$ and that $\mathcal{D}%
^{l}(\mathbb{Z}_{p})$ is a finite dimensional vector space spanned by the
basis $\left\{  \Omega\left(  p^{l}\left\vert x-i\right\vert _{p}\right)
\right\}  _{i\in G_{l}}$.

By identifying $\varphi\in\mathcal{D}^{l}(\mathbb{Z}_{p})$ with the column
vector $\left[  \varphi\left(  i\right)  \right]  _{i\in G_{l}}\in
\mathbb{R}^{\#G_{l}}$, we get that $\mathcal{D}^{l}(\mathbb{Z}_{p})$ is
isomorphic to $\mathbb{R}^{\#G_{l}}$ endowed with the norm $\left\Vert \left[
\varphi\left(  i\right)  \right]  _{i\in G_{l}^{N}}\right\Vert =\max_{i\in
G_{l}}\left\vert \varphi\left(  i\right)  \right\vert $. Furthermore,
\[
\mathcal{D}^{l}\hookrightarrow\mathcal{D}^{l+1}\hookrightarrow\mathcal{D}%
(\mathbb{Z}_{p}),
\]
where $\hookrightarrow$ denotes a continuous embedding.

The restriction of $E$ to the subspace $\mathcal{D}^{l}(\mathbb{Z}_{p})$ gives
a discretization of $E$ denoted as $E_{l}$. Indeed, by assuming that
\begin{align*}
\boldsymbol{v}\left(  x\right)   &  =%
%TCIMACRO{\tsum \limits_{i\in G_{l}}}%
%BeginExpansion
{\textstyle\sum\limits_{i\in G_{l}}}
%EndExpansion
\boldsymbol{v}\left(  i\right)  \Omega\left(  p^{l}\left\vert x-i\right\vert
_{p}\right)  \text{,}\\
\boldsymbol{h}\left(  x\right)   &  =%
%TCIMACRO{\tsum \limits_{i\in G_{l}}}%
%BeginExpansion
{\textstyle\sum\limits_{i\in G_{l}}}
%EndExpansion
\boldsymbol{h}\left(  i\right)  \Omega\left(  p^{l}\left\vert x-i\right\vert
_{p}\right)  ,
\end{align*}
we have
\begin{align*}
&
%TCIMACRO{\diint \limits_{\mathbb{Z}_{p}\times\mathbb{Z}_{p}}}%
%BeginExpansion
{\displaystyle\iint\limits_{\mathbb{Z}_{p}\times\mathbb{Z}_{p}}}
%EndExpansion
w\left(  x-y\right)  \Omega\left(  p^{l}\left\vert x-i\right\vert _{p}\right)
\Omega\left(  p^{l}\left\vert y-j\right\vert _{p}\right)  dxdy\\
&  =\iint\limits_{p^{l}\mathbb{Z}_{p}\times p^{l}\mathbb{Z}_{p}}%
w((i-j)+(x-y))dxdy,
\end{align*}
and by using that $a\left(  x\right)  $, $b(x)$, $c(x)$, $d(x)$ are test
functions supported in the unit ball, and taking $l$ sufficiently large, we
have%
\[
a\left(  x\right)  \Omega\left(  p^{l}\left\vert x-i\right\vert _{p}\right)
=a\left(  i\right)  \Omega\left(  p^{l}\left\vert x-i\right\vert _{p}\right)
\text{, }%
\]%
\[
b(x)\Omega\left(  p^{l}\left\vert x-i\right\vert _{p}\right)  =b(i)\Omega
\left(  p^{l}\left\vert x-i\right\vert _{p}\right)  ,
\]%
\[
c\left(  x\right)  \Omega\left(  p^{l}\left\vert x-i\right\vert _{p}\right)
=c\left(  i\right)  \Omega\left(  p^{l}\left\vert x-i\right\vert _{p}\right)
\text{,}%
\]%
\[
\text{ }d(x)\Omega\left(  p^{l}\left\vert x-i\right\vert _{p}\right)
=d(i)\Omega\left(  p^{l}\left\vert x-i\right\vert _{p}\right)  ,
\]
and consequently
\begin{gather*}
E_{l}\left(  \boldsymbol{v},\boldsymbol{h}\right)  =-%
%TCIMACRO{\dsum \limits_{\substack{i,\text{ }j\in G_{l}\\i\neq j}}}%
%BeginExpansion
{\displaystyle\sum\limits_{\substack{i,\text{ }j\in G_{l}\\i\neq j}}}
%EndExpansion
\boldsymbol{v}\left(  i\right)  \boldsymbol{h}\left(  j\right)  \left(  \text{
}\iint\limits_{p^{l}\mathbb{Z}_{p}\times p^{l}\mathbb{Z}_{p}}%
w(i-j+x-y)dxdy\right) \\
-p^{-l}%
%TCIMACRO{\dsum \limits_{i\in G_{l}}}%
%BeginExpansion
{\displaystyle\sum\limits_{i\in G_{l}}}
%EndExpansion
a\left(  i\right)  \boldsymbol{v}\left(  i\right)  -p^{-l}%
%TCIMACRO{\dsum \limits_{i\in G_{l}}}%
%BeginExpansion
{\displaystyle\sum\limits_{i\in G_{l}}}
%EndExpansion
b\left(  i\right)  \boldsymbol{h}\left(  i\right)  +\frac{ep^{-l}}{2}%
%TCIMACRO{\dsum \limits_{\boldsymbol{i}\in G_{l}}}%
%BeginExpansion
{\displaystyle\sum\limits_{\boldsymbol{i}\in G_{l}}}
%EndExpansion
\boldsymbol{v}^{2}\left(  i\right)  +\frac{ep^{-l}}{2}%
%TCIMACRO{\dsum \limits_{\boldsymbol{i}\in G_{l}}}%
%BeginExpansion
{\displaystyle\sum\limits_{\boldsymbol{i}\in G_{l}}}
%EndExpansion
\boldsymbol{h}^{2}\left(  i\right) \\
+\frac{p^{-l}}{2}%
%TCIMACRO{\dsum \limits_{i\in G_{l}}}%
%BeginExpansion
{\displaystyle\sum\limits_{i\in G_{l}}}
%EndExpansion
c\left(  i\right)  \boldsymbol{v}^{4}\left(  i\right)  +\frac{p^{-l}}{2}%
%TCIMACRO{\dsum \limits_{i\in G_{l}}}%
%BeginExpansion
{\displaystyle\sum\limits_{i\in G_{l}}}
%EndExpansion
d\left(  i\right)  \boldsymbol{h}^{4}\left(  i\right)  .
\end{gather*}

We take $v_{i}=\boldsymbol{v}\left(  i\right)  $, $h_{i}=\boldsymbol{h}\left(
i\right)  $,
\[
w_{i-j}=\iint\limits_{p^{l}\mathbb{Z}_{p}\times p^{l}\mathbb{Z}_{p}%
}w((i-j)+(x-y))dxdy,
\]
$a_{i}=p^{-l}a\left(  i\right)  $, $b_{i}=p^{-l}b\left(  i\right)  $,
$c_{i}=p^{-l}c\left(  i\right)  $, $d_{i}=p^{-l}d\left(  i\right)  $, for $i$,
$j\in G_{l}$, and $\boldsymbol{\theta}=\left\{  w_{ij},a_{i},b_{i},c_{i}%
,d_{i}\right\}  $. We also rescale $e$ to $ep^{l}$, then%
\begin{gather*}
E_{l}\left(  \boldsymbol{v},\boldsymbol{h};\boldsymbol{\theta}\right)
=-{\sum\limits_{i,\text{ }j\in G_{l}}}w_{i-j}v_{i}h_{j}-%
%TCIMACRO{\dsum \limits_{i\in G_{l}}}%
%BeginExpansion
{\displaystyle\sum\limits_{i\in G_{l}}}
%EndExpansion
a_{i}v_{i}-%
%TCIMACRO{\dsum \limits_{i\in G_{l}}}%
%BeginExpansion
{\displaystyle\sum\limits_{i\in G_{l}}}
%EndExpansion
b_{i}h_{i}\\
+\frac{e}{2}%
%TCIMACRO{\dsum \limits_{i\in G_{l}}}%
%BeginExpansion
{\displaystyle\sum\limits_{i\in G_{l}}}
%EndExpansion
v_{i}^{2}+\frac{e}{2}%
%TCIMACRO{\dsum \limits_{i\in G_{l}}}%
%BeginExpansion
{\displaystyle\sum\limits_{i\in G_{l}}}
%EndExpansion
h_{i}^{2}+%
%TCIMACRO{\dsum \limits_{i\in G_{l}}}%
%BeginExpansion
{\displaystyle\sum\limits_{i\in G_{l}}}
%EndExpansion
c_{i}v_{i}^{4}+%
%TCIMACRO{\dsum \limits_{i\in G_{l}}}%
%BeginExpansion
{\displaystyle\sum\limits_{i\in G_{l}}}
%EndExpansion
d_{i}h_{i}^{4}\text{.}%
\end{gather*}
We now recall that $G_{l}$ is an additive group,\ then%
\[%
%TCIMACRO{\dsum \limits_{i,\text{ }j\in G_{l}}}%
%BeginExpansion
{\displaystyle\sum\limits_{i,\text{ }j\in G_{l}}}
%EndExpansion
w_{i-j}v_{i}h_{j}=%
%TCIMACRO{\dsum \limits_{j\in G_{l}}}%
%BeginExpansion
{\displaystyle\sum\limits_{j\in G_{l}}}
%EndExpansion%
%TCIMACRO{\dsum \limits_{k\in G_{l}}}%
%BeginExpansion
{\displaystyle\sum\limits_{k\in G_{l}}}
%EndExpansion
w_{k}v_{j+k}h_{j},
\]
and consequently%
\begin{gather*}
E_{l}\left(  \boldsymbol{v},\boldsymbol{h};\boldsymbol{\theta}\right)  =-%
%TCIMACRO{\dsum \limits_{j\in G_{l}}}%
%BeginExpansion
{\displaystyle\sum\limits_{j\in G_{l}}}
%EndExpansion%
%TCIMACRO{\dsum \limits_{k\in G_{l}}}%
%BeginExpansion
{\displaystyle\sum\limits_{k\in G_{l}}}
%EndExpansion
w_{k}v_{j+k}h_{j}-%
%TCIMACRO{\dsum \limits_{j\in G_{l}}}%
%BeginExpansion
{\displaystyle\sum\limits_{j\in G_{l}}}
%EndExpansion
a_{j}v_{j}-%
%TCIMACRO{\dsum \limits_{j\in G_{l}}}%
%BeginExpansion
{\displaystyle\sum\limits_{j\in G_{l}}}
%EndExpansion
b_{j}h_{j}\\
+\frac{e}{2}%
%TCIMACRO{\dsum \limits_{j\in G_{l}^{N}}}%
%BeginExpansion
{\displaystyle\sum\limits_{j\in G_{l}^{N}}}
%EndExpansion
v_{j}^{2}+\frac{e}{2}%
%TCIMACRO{\dsum \limits_{j\in G_{l}}}%
%BeginExpansion
{\displaystyle\sum\limits_{j\in G_{l}}}
%EndExpansion
h_{j}^{2}+%
%TCIMACRO{\dsum \limits_{j\in G_{l}}}%
%BeginExpansion
{\displaystyle\sum\limits_{j\in G_{l}}}
%EndExpansion
c_{j}v_{j}^{4}+%
%TCIMACRO{\dsum \limits_{j\in G_{l}}}%
%BeginExpansion
{\displaystyle\sum\limits_{j\in G_{l}}}
%EndExpansion
d_{j}h_{j}^{4}\text{.}%
\end{gather*}

\section{\label{Section8}Appendix C: Some probability distributions}

%\subsection{Boltzmann probability distributions}

From now on, we assume that visible and hidden fields are binary variables.
However, most of our mathematical formulation is valid under the assumption that
visible and hidden fields are discrete variables. We set
\[
\boldsymbol{V}=\{V_{1},V_{2},\ldots,V_{N}\},\quad\boldsymbol{H}=\{H_{1}%
,H_{2},\ldots,H_{N}\},
\]
$N=p^{l}$, to be the respective visible and hidden random variable sets. The
random variables $(\boldsymbol{V},\boldsymbol{H})$ take values
$(\boldsymbol{v},\boldsymbol{h})\in\{0,1\}^{2N}$. For the sake of simplicity, we will identify the random vector $\boldsymbol{V}$ with $\boldsymbol{v}$, and the random vector $\boldsymbol{H}$ with $\boldsymbol{h}$. We identify $G_{l}$ with the
set of branches (vertices at the top level) of a rooted tree with $l+1$ levels
and $p^{l}$ branches. Attached to each branch $i\in G_{l}$ there are two are
two states: $v_{i}$, $h_{i}$. With this notation, $\boldsymbol{v}=\left[
v_{i}\right]  _{i\in G_{l}}$ is a realization of the\textit{\ }visible field,
and $\boldsymbol{h}=\left[  h_{i}\right]  _{i\in G_{l}}$ is a realization of
the hidden field. 
%The \ variables $v_{i}$, $v_{j}$ given $\boldsymbol{h}$,
%respectively $h_{i}$, $h_{j}$ given $\boldsymbol{v}$, are independent for
%$i\neq j$.
The joint distribution of the random vectors $(\boldsymbol{v},\boldsymbol{h}%
)$\ is given by the following Boltzmann probability distribution:%
\begin{equation}
\boldsymbol{P}_{l}(\boldsymbol{v},\boldsymbol{h};\boldsymbol{\theta}%
)=\frac{\exp\left(  -E_{l}\left(  (\boldsymbol{v},\boldsymbol{h}%
;\boldsymbol{\theta})\right)  \right)  }{Z_{l}}, \label{Joint_Distribution}%
\end{equation}
where%
\[
Z_{l}={\sum\limits_{\boldsymbol{v},\boldsymbol{h}}}\exp\left(  -E_{l}\left(
\boldsymbol{v},\boldsymbol{h};\boldsymbol{\theta}\right)  \right),
\]
and $E_{l}$ is defined in \Cref{energy-discretization}

%\subsection{The marginal probability distributions}

By using the joint distribution of the visible field and the hidden field
(\ref{Joint_Distribution}), we compute the marginal probability distributions
as follows:%
\[
\boldsymbol{P}_{l}(\boldsymbol{v};\boldsymbol{\theta})=%
%TCIMACRO{\dsum \limits_{\boldsymbol{h}}}%
%BeginExpansion
{\displaystyle\sum\limits_{\boldsymbol{h}}}
%EndExpansion
\boldsymbol{P}_{l}(\boldsymbol{v},\boldsymbol{h};\boldsymbol{\theta}%
)=\frac{{\sum\limits_{\boldsymbol{h}}}\exp\left(  -E_{l}\left(  \boldsymbol{v}%
,\boldsymbol{h};\boldsymbol{\theta}\right)  \right)  }{%
%TCIMACRO{\dsum \limits_{\boldsymbol{v},\boldsymbol{h}}}%
%BeginExpansion
{\displaystyle\sum\limits_{\boldsymbol{v},\boldsymbol{h}}}
%EndExpansion
\exp\left(  -E_{l}\left(  \boldsymbol{v},\boldsymbol{h};\boldsymbol{\theta
}\right)  \right)  },
\]%
\[
\boldsymbol{P}_{l}(\boldsymbol{h};\boldsymbol{\theta})=%
%TCIMACRO{\dsum \limits_{\boldsymbol{v}}}%
%BeginExpansion
{\displaystyle\sum\limits_{\boldsymbol{v}}}
%EndExpansion
\boldsymbol{P}_{l}(\boldsymbol{v},\boldsymbol{h};\boldsymbol{\theta})=\frac{%
%TCIMACRO{\dsum \limits_{\boldsymbol{v}}}%
%BeginExpansion
{\displaystyle\sum\limits_{\boldsymbol{v}}}
%EndExpansion
\exp\left(  -E_{l}\left(  \boldsymbol{v},\boldsymbol{h};\boldsymbol{\theta
}\right)  \right)  }{%
%TCIMACRO{\dsum \limits_{\boldsymbol{v},\boldsymbol{h}}}%
%BeginExpansion
{\displaystyle\sum\limits_{\boldsymbol{v},\boldsymbol{h}}}
%EndExpansion
\exp\left(  -E_{l}\left(  \boldsymbol{v},\boldsymbol{h};\boldsymbol{\theta
}\right)  \right)  }.
\]

%\subsection{The conditional probability distributions}\label{subsec:independence}

Since we are assuming that $\boldsymbol{v},\boldsymbol{h}$ are binary, the
energy functional takes the form
\[
E_{l}(\boldsymbol{v},\boldsymbol{h};\boldsymbol{\theta})=-\sum_{k\in G_{l}%
}\sum_{j\in G_{l}}w_{k}v_{j+k}h_{j}-\sum_{j\in G_{l}}a_{j}v_{j}-\sum_{j\in
G_{l}}b_{j}h_{j}.
\]
The classical RBM has the advantage of independence between the visible units
as well as the hidden units. The $p$-adic BM shares the same advantage, i.e.,
by fixing the hidden field $\boldsymbol{h}$, the random variables $v_{i}$,
$i\in G_{l}$, become independent. An analog assertion is valid if we fix the
visible field $\boldsymbol{v}$. More precisely, the conditional probability
distributions satisfy%
\[
\boldsymbol{P}_{l}(\boldsymbol{v}\mid\boldsymbol{h};\boldsymbol{\theta
})={\prod\limits_{j\in G_{l}}}\boldsymbol{P}_{l}\left(  v_{j}\mid
\boldsymbol{h};\boldsymbol{\theta}\right), \quad
\mbox{and} \quad
\boldsymbol{P}_{l}(\boldsymbol{h}\mid\boldsymbol{v};\boldsymbol{\theta})=%
%TCIMACRO{\dprod \limits_{j\in G_{l}}}%
%BeginExpansion
{\displaystyle\prod\limits_{j\in G_{l}}}
%EndExpansion
\boldsymbol{P}_{l}\left(  h_{j}\mid\boldsymbol{v};\boldsymbol{\theta}\right).
\]
Indeed, by direct computation, we have

\begin{align*}
&\boldsymbol{P}_{l}(\boldsymbol{v}\mid\boldsymbol{h};\boldsymbol{\theta}%
)
=\frac{\boldsymbol{P}_{l}(\boldsymbol{v},\boldsymbol{h};\boldsymbol{\theta}%
)}{\boldsymbol{P}_{l}(\boldsymbol{h};\boldsymbol{\theta})}
\\
=&\frac{\exp\left(  -E_{l}\left(  \boldsymbol{v},\boldsymbol{h}%
;\boldsymbol{\theta}\right)  \right)  }{
{\displaystyle\sum\limits_{\boldsymbol{v}}}
\exp\left(  -E_{l}\left(  \boldsymbol{v},\boldsymbol{h};\boldsymbol{\theta
}\right)  \right)  }
=\frac{
{\displaystyle\prod\limits_{j\in G_{l}}}
\exp\left(
{\displaystyle\sum\limits_{k\in G_{l}}}
w_{k}v_{j}h_{j-k}+a_{j}v_{j}\right)  }{
{\displaystyle\sum\limits_{\boldsymbol{v}}}
{\displaystyle\prod\limits_{j\in G_{l}}}
\exp\left(
{\displaystyle\sum\limits_{k\in G_{l}}}
w_{k}v_{j}h_{j-k}+a_{j}v_{j}\right)  }\\
=&
{\displaystyle\prod\limits_{j\in G_{l}}}
\frac{\exp\left(  v_{j}
{\displaystyle\sum\limits_{k\in G_{l}}}
w_{k}h_{j-k}+a_{j}v_{j}\right)  }{%
{\displaystyle\sum\limits_{v_{j}}}
\exp\left(  v_{j}
{\displaystyle\sum\limits_{k\in G_{l}}}
w_{k}h_{j-k}+a_{j}v_{j}\right)  }
=
{\displaystyle\prod\limits_{j\in G_{l}}}
\frac{\boldsymbol{P}_{l}\left(  v_{j},\boldsymbol{h};\boldsymbol{\theta
}\right)  }{
{\displaystyle\sum\limits_{v_{j}}}
\boldsymbol{P}_{l}\left(  v_{j},\boldsymbol{h};\boldsymbol{\theta}\right)  }=
{\displaystyle\prod\limits_{j\in G_{l}}}
\boldsymbol{P}_{l}\left(  v_{j}\mid\boldsymbol{h};\boldsymbol{\theta}\right).
\end{align*}

Similarly, we can prove that
\begin{equation}
\boldsymbol{P}_{l}(\boldsymbol{h}\mid\boldsymbol{v};\boldsymbol{\theta
})={\prod\limits_{j\in G_{l}}}\boldsymbol{P}_{l}\left(  h_{j}\mid
\boldsymbol{v};\boldsymbol{\theta}\right)  . \label{Formula_p_h_v}%
\end{equation}

\subsection{Gradient of the Log-likelihood}

The log-likelihood giving a single visible state $\boldsymbol{v}$ is given
\[
\ln\boldsymbol{P}_{l}(\boldsymbol{v}|\boldsymbol{\theta})=\ln\dfrac{1}{Z}%
\sum_{\boldsymbol{h}}e^{-E(\boldsymbol{v},\boldsymbol{h})}=\ln\sum
_{\boldsymbol{h}}e^{-E(\boldsymbol{v},\boldsymbol{h})}-\ln\sum_{\boldsymbol{h}%
,\boldsymbol{v}}e^{-E(\boldsymbol{v},\boldsymbol{h})}.
\]
Taking the derivative with respect to the parameters gives the following
mean-like representation:%
\begin{equation}\label{log-likelihood}
\begin{split}
&\dfrac{\partial}{\partial\boldsymbol{\theta}}\ln\boldsymbol{P}_{l}%
(\boldsymbol{v}|\boldsymbol{\theta})=\dfrac{\partial}{\partial
\boldsymbol{\theta}}\ln\sum_{\boldsymbol{h}}e^{-E(\boldsymbol{v}%
,\boldsymbol{h})}-\dfrac{\partial}{\partial\boldsymbol{\theta}}\ln
\sum_{\boldsymbol{h},\boldsymbol{v}}e^{-E(\boldsymbol{v},\boldsymbol{h}% 
)}\\
=&
-\dfrac{1}{\sum\limits_{\boldsymbol{h}}e^{-E(\boldsymbol{v},\boldsymbol{h})}}%
\sum_{\boldsymbol{h}}e^{-E(\boldsymbol{v},\boldsymbol{h})}\dfrac{\partial
}{\partial\boldsymbol{\theta}}E(\boldsymbol{v},\boldsymbol{h})
+\dfrac{1}%
{\sum\limits_{\boldsymbol{v},\boldsymbol{h}}e^{-E(\boldsymbol{v},\boldsymbol{h})}%
}\sum_{\boldsymbol{v},\boldsymbol{h}}e^{-E(\boldsymbol{v},\boldsymbol{h}%
)}\dfrac{\partial}{\partial\boldsymbol{\theta}}E(\boldsymbol{v},\boldsymbol{h}%
)\\
=&-\sum_{\boldsymbol{h}}\boldsymbol{P}_{l}(\boldsymbol{h}|\boldsymbol{v}%
)\dfrac{\partial}{\partial\boldsymbol{\theta}}E(\boldsymbol{v},\boldsymbol{h}%
)
+\sum_{\boldsymbol{v},\boldsymbol{h}}\boldsymbol{P}_{l}(\boldsymbol{v}%
,\boldsymbol{h})\dfrac{\partial}{\partial\boldsymbol{\theta}}E(\boldsymbol{v}%
,\boldsymbol{h}).
\end{split}
\end{equation}
In the case of multiple visible states $S=\{\boldsymbol{v}_{1},\boldsymbol{v}%
_{2},\cdots,\boldsymbol{v}_{s}\}$, the log-likelihood is defined in the
average sense, i.e., $\dfrac{1}{s}\sum\limits_{\boldsymbol{v}\in S}\ln\boldsymbol{P}%
_{l}(\boldsymbol{v}|\boldsymbol{\theta})$.

Taking the derivative with respect to $w_{k}$ gives
\begin{equation}\label{derivative-wk}
\begin{split}
	&\dfrac{\partial}{\partial w_k} \ln \boldsymbol{P}_{l}(\bv| \btheta)  	
	%-  \sum_{\bh} \boldsymbol{P}_{l}(\bh|\bv)  \dfrac{\partial}{\partial w_k} E(\bv, \bh)
	%+\sum_{\bv, \bh} \boldsymbol{P}_{l}(\bh,\bv)  \dfrac{\partial}{\partial w_k} E(\bv, \bh)\\
	=\sum_{\bh} \left( \boldsymbol{P}_{l}(\bh|\bv) \sum_{j \in G_l} v_{j+k} h_j \right)
	-\sum_{\bv, \bh}  \left(  \boldsymbol{P}_{l}(\bh,\bv)  \sum_{j \in G_l} v_{j+k} h_j \right)\\
	=&\sum_{\bh} \left( \boldsymbol{P}_{l}(\bh|\bv) \sum_{j \in G_l} v_{j+k} h_j \right)
	-\sum_{\bv} \boldsymbol{P}_{l}(\bv) \left(  \sum_{\bh} \left(\boldsymbol{P}_{l}(\bh|\bv)   \sum_{j \in G_l} v_{j+k} h_j \right) \right).
	\end{split}
\end{equation}
Now, let $\mathbb{I}$ be the ordered indexes of all the positive states in $\bv$.
\begin{equation}\label{derivative-wk-1}
\begin{split}
	&\sum_{\bh} \left( \boldsymbol{P}_l(\bh|\bv) \sum_{i \in G_l} v_{i+k} h_i \right) = 
	\sum_{\bh}  \boldsymbol{P}_l(\bh|\bv) \left(\sum_{i \in \mathbb{I}} h_{i-k} \right) \\
	=&
	\sum_{\bh_{{\mathbb{I}}}} \sum_{\bh_{G_{l} \setminus \mathbb{I}}} 
	\prod_{i \in\ \mathbb{I}}\boldsymbol{P}_l(h_{i-k}|\bv) 
	\prod_{i \in G_l \setminus\ \mathbb{I}}\boldsymbol{P}_l(h_{i-k}|\bv)
	 \left(\sum_{i \in\ \mathbb{I}} h_{i-k} \right)  \\
	 =&
	\left(\sum_{\bh_{{\mathbb{I}}}} 
	\prod_{i \in\ \mathbb{I}}\boldsymbol{P}_l(h_{i-k}|\bv)  \left(\sum_{i \in\ \mathbb{I}} h_{i-k} \right)  \right)
	\underbrace{
	\left(\sum_{\bh_{G_l \setminus\ \mathbb{I}}} \prod_{i \in G_l \setminus\ \mathbb{I}}\boldsymbol{P}_l(h_{i-k}|\bv) \right)}_{=1}\\
	=&
	\sum_{\bh_{{\mathbb{I}}}} \left(
	 \displaystyle\prod_{i \in {{\mathbb{I}}}}
	\boldsymbol{P}_l(h_{i-k}|\bv)   \sum_{i \in {{\mathbb{I}}}} h_{i-k} \right)
	\\
	=& \sum_{{i \in\ \mathbb{I}}} \boldsymbol{P}_l(h_{i-k} =1|\bv) 
	=  \sum_{i \in G_l}\boldsymbol{P}_l(h_{i-k} =1| \bv) v_i.
	\end{split}
\end{equation}

Combing \cref{derivative-wk} and \cref{derivative-wk-1} gives
\begin{gather}
\frac{\partial\log\boldsymbol{P}_{l}(\boldsymbol{v}|\boldsymbol{\theta}%
)}{w_{k}}
=
\sum_{i \in G_l}\boldsymbol{P}_l(h_{i-k} =1|\bv) v_i
-
\sum_{\boldsymbol{v}}\boldsymbol{P}%
_{l}(\boldsymbol{v})
\left(\sum_{i \in G_l}\boldsymbol{P}_l(h_{i-k} =1|\bv) v_i \right).%
\label{Eq- Log-likelihood}
%=\sum_{j\in G_{l}}\boldsymbol{P}_{l}(h_{j}=1|\boldsymbol{v})v_{j+k}%
%-\sum_{\boldsymbol{v}}\boldsymbol{P}_{l}(\boldsymbol{v})\boldsymbol{P}%
%_{l}(h_{j}=1|\boldsymbol{v})v_{j+k}.\nonumber
\end{gather}

Note that the above formula is different from the classical RBM, \cite[Formula
29]{Fischer-Igel}.

We derive the derivatives with respect to $a_{j}$ and $b_{j}$ similarly as in the classical RBM:
\begin{gather}
\dfrac{\partial}{\partial a_{j}}\ln\boldsymbol{P}_{l}(\boldsymbol{v}%
|\boldsymbol{\theta})=-\sum_{\boldsymbol{h}}\boldsymbol{P}_{l}(\boldsymbol{h}%
|\boldsymbol{v})\dfrac{\partial}{\partial a_{j}}E_{l}(\boldsymbol{v}%
,\boldsymbol{h})+\sum_{\boldsymbol{v},\boldsymbol{h}}\boldsymbol{P}%
_{l}(\boldsymbol{h},\boldsymbol{v})\dfrac{\partial}{\partial a_{j}}%
E_{l}(\boldsymbol{v},\boldsymbol{h}) \nonumber\\
=\sum_{\boldsymbol{h}}\boldsymbol{P}_{l}(\boldsymbol{h}|\boldsymbol{v}%
)v_{j}-\sum_{\boldsymbol{v},\boldsymbol{h}}\boldsymbol{P}_{l}(\boldsymbol{h}%
,\boldsymbol{v})v_{j}
=v_{j}-\sum_{\boldsymbol{v}}\boldsymbol{P}_{l}%
(\boldsymbol{v})v_{j}, \label{derivative-a}
\end{gather}
and%
\begin{gather}
\dfrac{\partial}{\partial b_{j}}\ln\boldsymbol{P}_{l}(\boldsymbol{v}%
|\boldsymbol{\theta})=-\sum_{\boldsymbol{h}}\boldsymbol{P}_{l}(\boldsymbol{h}%
|\boldsymbol{v})\dfrac{\partial}{\partial b_{j}}E_{l}(\boldsymbol{v}%
,\boldsymbol{h})+\sum_{\boldsymbol{v},\boldsymbol{h}}\boldsymbol{P}%
_{l}(\boldsymbol{h},\boldsymbol{v})\dfrac{\partial}{\partial b_{j}}%
E_{l}(\boldsymbol{v},\boldsymbol{h}) \nonumber\\
=\sum_{\boldsymbol{h}}\boldsymbol{P}_{l}(\boldsymbol{h}|\boldsymbol{v}%
)h_{j}-\sum_{\boldsymbol{v},\boldsymbol{h}}\boldsymbol{P}_{l}(\boldsymbol{h}%
,\boldsymbol{v})h_{j}=\boldsymbol{P}_{l}(h_{j}=1|\boldsymbol{v})-\sum
_{\boldsymbol{v}}\boldsymbol{P}_{l}(h_{j}=1|\boldsymbol{v}). \label{derivative-b}
\end{gather}

\subsection{Contrastive divergence learning}\label{Sec:CD_learning}

As in the classical case, the exact computation of the gradient of the
log-likelihood involves an exponential number of terms, see
(\ref{Eq- Log-likelihood}). We adopt the contrastive divergence (CD) method,
introduced by Hinton \cite{Hinton}, to the $p$-adic case to approximate the
minimization of the gradient of the log-likelihood.

The approximation of \cref{Eq- Log-likelihood} --\cref{derivative-b} using the contrastive convergence method 
 can be represented as follows:%
\begin{equation}
\begin{split}
&\frac{\partial\log\boldsymbol{P}_{l}(\boldsymbol{v}|\boldsymbol{\theta}%
)}{w_{k}} \approx \sum_{{i\in G_{l}}}\boldsymbol{P}_{l}(h_{i-k}=1|\boldsymbol{v}%
^{(0)})v_{i}-\sum_{{i\in G_{l}}}\boldsymbol{P}_{l}(h_{i-k}=1|\boldsymbol{v}%
^{(m)})v_{i}^{(m)},\\
&\frac{\partial\log\boldsymbol{P}_{l}(\boldsymbol{v}|\boldsymbol{\theta}%
)}{a_{j}} \approx
v_{j}^{(0)}-v_{j}^{(m)},\\
&\frac{\partial\log\boldsymbol{P}_{l}(\boldsymbol{v}|\boldsymbol{\theta}%
)}{b_{j}} \approx
\boldsymbol{P}_{l}(h_{j}=1|\boldsymbol{v}^{(0)})-
\boldsymbol{P}_{l}(h_{j}=1|\boldsymbol{v}^{(m)}), \label{derivative-b}
\end{split}
\end{equation}
where $m$ is a pre-determined positive integer,
 $\bv^{(0)}$ is a training example and 
$\bv^{(m)}$ is a sample of the Gibbs chain  after  $m$ steps.
More precisely,
%The computation of the discrete partition function is an intractable problem,
%since $\frac{\partial_{t}\boldsymbol{P}_{l}(\boldsymbol{v}|\boldsymbol{\theta
%})}{\partial w_{k}}$ depends on the discrete partition, we use the contrastive
%diverence method to approximate $\frac{\partial_{t}\boldsymbol{P}%
%_{l}(\boldsymbol{v}|\boldsymbol{\theta})}{\partial w_{k}}$ by $CD_{k}^{(m)}$.
we implement the Gibbs sampling in the following way. First, 
%define
%$\boldsymbol{v}^{(0)}$ as $\boldsymbol{v}$ and 
we obtain a sample of $\boldsymbol{h}^{(0)}$ 
using the conditional distribution $P_{l}(\boldsymbol{h}|\boldsymbol{v^{(0)}})$.
Then, we obtain a sample of
$\boldsymbol{v}^{(1)}$ using $P_{l}(\boldsymbol{v}|\boldsymbol{h}^{(0)})$. We
repeat this process until we get $\boldsymbol{v}^{(m)}$.

The following formulas are utilized in the calculation. Let
$\boldsymbol{h}_{-i}$ denotes the state of all hidden units expect for the
$i$-th one:%

\begin{gather*}
\boldsymbol{P}_{l}(h_{i}=1|\boldsymbol{v})
=\boldsymbol{P}_{l}(h_{i}=1|\boldsymbol{h}_{-i},\boldsymbol{v})
=\dfrac{\boldsymbol{P}_{l}(h_{i}=1,\boldsymbol{h}_{-i},\boldsymbol{v})}{\boldsymbol{P}_{l}%
(\boldsymbol{h}_{-i},\boldsymbol{v})}\\
=\dfrac{\boldsymbol{P}_{l}(h_{i}=1,\boldsymbol{h}_{-i},\boldsymbol{v}%
)}{\boldsymbol{P}_{l}(h_{i}=1,\boldsymbol{h}_{-i},\boldsymbol{v}%
)+\boldsymbol{P}_{l}(h_{i}=0,\boldsymbol{h}_{-i},\boldsymbol{v})}=\dfrac
{1}{1+\dfrac{\boldsymbol{P}_{l}(h_{i}=0,\boldsymbol{h}_{-i},\boldsymbol{v}%
)}{\boldsymbol{P}_{l}(h_{i}=1,\boldsymbol{h}_{-i},\boldsymbol{v})}}\\
=
\dfrac{1}{1+\dfrac{1}{\exp\left(\sum\limits_{k\in G_{l}}w_{k}v_{i+k}+b_{i}\right)}}%
=:\sigma\left(  {\sum_{k\in G_{l}}w_{k}v_{i+k}+b_{i})}\right)  .
\end{gather*}
Similarly, we have%
\begin{equation}
\begin{split}
&\boldsymbol{P}_{l}(h_{i}=0|\boldsymbol{v})
=\sigma\left(  -{\sum_{k\in G_{l}}w_{k}v_{i+k}-b_{i}}\right),\\
&\boldsymbol{P}_{l}(v_i = 1|\bh) =
	 \sigma \left( \sum_{k \in G_l} w_{i-k} h_{k} + a_i \right),\\
&\boldsymbol{P}_{l}(v_i = 0|\bh)  =
	 \sigma \left(-\sum_{k \in G_l} w_{i-k} h_{k} - a_i\right).
\end{split}
\end{equation}

\end{document}